# A Path towards Thresholdless Colloidal Quantum Dot Lasers

# by Solving Decades of Mythology on Optical Gain

Davide Zenatti and Patanjali Kambhampati*

Department of Chemistry, McGill University, Montreal, Canada

*pat.kambhampati@mcgill.ca

# Abstract

The semiconductor quantum dot (CQD) was first conceived in the 1980s as offering potential for future lasers. Following high quality solution phase synthesis of colloidal CQD (CCQD) in 1993, optical gain was first demonstrated in 2000 via stimulated emission (SE) measurements. Decades of phenomenology have given rise to a Standard Model of optical gain in CCQD based upon two assumptions: a biexciton is required to achieve optical gain, and short biexciton lifetimes limit the efficient development of amplified spontaneous emission (ASE). The Standard Model predicts the solution to efficient ASE is slowing Auger recombination, a task which now consumes the CCQD field. Here, we show that the Standard Model is physically incorrect, leading to misdirected materials development. Inspection of the phenomenology reveals a Simple Model which uniquely reproduces all observed phenomena. The Simple Model provides the physical foundation for developing CCQD lasers with quantum leaps in performance, including thresholdless gain.

**Keywords:** colloidal quantum dot, nanocrystal, optical gain, stimulated emission, amplified spontaneous emission, biexciton

# 1. Towards the Standard Model of Optical Gain in Quantum Dots

## 1.1. A Box for Electrons

The semiconductor quantum dot (CQD) was proposed in the 1980s as a structure that would confine electrons to produce new laser gain media[1]. Based upon existing semiconductor lasers based upon bulk semiconductors, nanoscale versions were subsequently created such as the quantum well. Subsequently the one-dimensional confinement of quantum wells was extended to three-dimensional confinement in quantum dots. In all these cases, the growth of the material was through vacuum based epitaxial methods. Shortly after, colloidal quantum dots (CCQD) were developed in which growth took place via solution phase chemistry[1]. Based upon the solution processibility of these CCQD, they offer attractive features such as the ability to make strongly confined CQD and integration into small form factor devices[2-7].

The idea driving the development of quantum confined semiconductors and ultimately the CQD is illustrated in **Fig1**. Shown in **Fig1a** are illustrations of semiconductors in the bulk phase, the quantum well phase, and the quantum dot phase. Shown in **Fig1b** are schematic representations of the density of electronic states in each system. Shown in **Fig1c** are illustrations of a realistic colloidal CQD, and a biexcitonic state in a CQD. In a bulk semiconductor the density of states is a continuous function with an $E^{1/2}$ functional form. Formation of a quantum well results in one dimensional quantum confinement giving rise to steps in the density of states. Formation of a CQD results in delta functions for each transition between states. It is this formation of delta function type features that is especially attractive to laser development. Such narrow spectral lines and discrete transitions enable lower gain thresholds and improved temperature performance[8].

Focusing on CCQD development, the formation of these structures was made possible by both solution phase synthesis and molten glass formation[1]. These initial CQD in the 1980s were not of sufficiently high quality to be useful for optoelectronic applications. An enabling step was made by the first synthesis of high quality Cd chalcogenide CQD by Bawendi in 1993[9]. With that synthetic method, it was now possible to produce CdSe CQD with narrow size distributions and relatively high photoluminescence (PL). The development of shelling by Guyot-Sionnest in 1997[10] further advanced these CQD for optical gain by increasing the PL quantum yield. Having achieved high quality synthesis of CdSe and related CQD in the 1990s, the next step was to test the early laser ideas and see optical gain. Once observed, the gain performance can then be quantified, understood theoretically, and realized via synthesis.

Following the first observation of optical gain in CdSe CQD by Klimov and Bawendi[11] there have been many verifications to generalize the process to a wide variety of CQD as discussed in recent reveiews[2, 3, 6, 12, 13]. After 25 years of development, there has been much progress towards the development of CQD lasers[14-20], beyond merely seeing the fundamental effect optical gain. Based upon decades of work one might think that the physics is well solved, and what remains is materials optimization based upon targets created by the physical models of optical gain in CQD.

Indeed, there is a Standard Model [2-4, 6, 12] for optical gain in these CQD that has been well validated in many experiments. The Standard Model invokes several assumptions: 1) the CQD is best described as a two-level system (2LS), 2) biexciton formation is required to generate optical gain as seen in Stimulated Emission (SE) measurements, 3) biexcitons undergo rapid recombination due to efficient quantum confinement enhanced Auger processes, 4) the fast biexciton recombination prevents the buildup of optical gain into Amplified Spontaneous Emission (ASE) which is the first step towards a laser, 5) the roadmap for the CQD materials researchers is to design new CQD with slow biexciton recombination.

Here, we propose that the prevailing Standard Model of CQD optical gain incorrectly assigns the optical gain efficiency to biexciton Auger recombination timescale. We show that the CQD lasing system cannot be a 2LS as proposed in the Standard Model. Instead, it must be at least a 3LS and may be engineered to be a 4LS, representing the ideal lasing system. Realization of a 4LS in a CQD would be thresholdless optical gain. The matching between biexciton lifetimes and ASE thresholds and amplitudes we reveal to be merely a fortuitous result of experimental design. In the place of the empirically derived Standard Model of optical gain in CQD, we propose a Simple Model which correctly classifies CQD into 3 / 4 / 5 level systems. Finding the CQDs that make the transition from a 3LS to a 4+LS, we show is the key step in the path towards optical gain by creating SE at low thresholds and with high cross sections. Slowing biexciton recombination has no influence on SE thresholds. The only reason it has an influence on ASE thresholds and cross sections arises from artifacts of experimental design and no fundamental physical process. This Simple Model of optical gain in CQD can explain all observed phenomena. It moreover enables predictions of new phenomena. These new phenomena, e.g. thresholdless gain, should be targets for future materials design to physically realize the proposals.

## 1.2. The biexciton and Optical Gain in Quantum Dots

The key idea of the Standard Model of optical gain in CQD is that biexciton formation is required to develop SE. **Fig1c** presents an overview of excitons and biexcitons in CQD. **Fig1c** schematically illustrates a CCQD (top), e.g. CdSe with a diameter equal to the Bohr length thus providing a confining potential. Shown to scale is a typical ligand which passivates the surface of these CQD. Notably the real CQD is not a sphere, but as atomistic details, including surfaces with terraces, ledges, and kinks [21-29]. These atomistic details become important in evaluating the performance of real CQD, and any additions to the initial models proposed here. Also shown (bottom) is a schematic illustration of one of many possible biexcitons in a CQD [30]. Shown here is a schematic illustration of an excited state biexciton comprised of an S type exciton and a P type exciton. This illustration of an excided biexciton state immediately raises questions on which biexcitons are important for optical gain.

The idea driving the Standard Model is shown in **Fig2a**, reproduced from Klimov[2]. The excitonic level structure arises from the conduction band (CB) and valence band (VB) being quantized into levels. The Standard Model assumes that the CQD is best described as a 2LS as shown. With this band edge exciton, there is a two-fold degeneracy for a 1S state in CdSe CQD. This band edge exciton is eight-fold degenerate in PbSe CQD based upon differences in the basis band structures. Focusing on CdSe CQD as the model system, an unexcited CQD consists of two ground state excitons. This system can absorb one or two photons for pulsed excitation. A singly excited CQD in this picture is optically transparent. A doubly excited CQD enables SE.

**Fib2b** further illustrates the exciton and biexciton level structure and how it relates to pumping and recombination rate constants. Notably the SE is now dependent on the biexciton recombination rate constant. This idea drives the conclusion that slower Auger based biexciton recombination would be beneficial for the development of SE and ultimately ASE. The problem with this picture along with the problems throughout the Standard Model is that optical gain in CQD is performed with pulsed excitation rather than continuous wave. Hence these kinetic rate equations are not relevant, and pulsed equations should be used as we discuss in detail below.

The proposal that the CQD is best described as an excitonic 2LS can immediately be dismissed upon examination of the simplest linear spectroscopy. Colloidal CQDs have a Stokes shift separating absorption into the band edge exciton from emission from the band edge exciton [9, 31-37]. This Stoke shift is not insignificant, being of similar magnitude as the linewidths (50 – 100 meV at 300K).

The Stokes shift was originally rationalized by the idea of an excitonic fine structure [31-33]. But the fine structure cannot describe the total Stokes shift. Hence phonon progressions were next included in the rationalization of the Stokes shifts [38-41]. Given the existence of a Stokes shift the CQD cannot be a 2LS as a class of lasing system. The presence of excitonic fine structure separating the absorbing state from the emitting state implies that the CQD must be at least at 3LS. Based upon the ability to control exciton-phonon couplings and phonon spectra, the CQD can transition from a 3LS to a 4+LS. This transition between classes of lasing systems is a key idea as the 2LS does not lase, the 3lS does lase at some finite threshold, and the 4LS has thresholdless gain in theory.

The fine structure to excitons remains under investigation to obtain an atomistic picture in experiment by Coherent Multi-Dimensional Spectroscopy [22, 42] (CMDS) and in theory by Empirical Pseudo-Potential Methods (EPM) [21-23, 26, 28, 29]. In addition to the coarse and fine electronic structure of the singe exciton [32, 33, 43, 44, 45], there is evidence for excited states of biexcitons [30, 38, 46-52]. With the experimental observation of an excitonic Stoke shift and a biexcitonic spectrum, combined with the theoretical explanation in terms of coarse and fine structure, it is impossible that the colloidal CQD can be conceived of as a two-level lasing system.

## 1.3. Slowing Biexciton Recombination Appears to Lower ASE Thresholds

Given the starting assumption of the Standard Model that a biexciton is required to generate optical gain in a 2LS, the main experimental aim has been to measure biexciton lifetimes, and the main materials development goal has been to lengthen these lifetimes. By slowing Auger recombination of biexcitons, the ASE thresholds were argued to be lowered and the amplitudes increased.

**Fig2c-d** shows the results on CCQD biexcitonic figures of merit, adapted from Klimov [2, 3, 12, 13]. **Fig2c** shows the biexciton non-radiative recombination time constant, $\tau_{XX}$, vs volume. included in this plot are results we obtained on CdSe[53] and CsPbBr$_3$[54, 55] CQD using improved measurement techniques, along with simultaneous measurements on the same improved spectrometer.

In all prior works, time-resolved photoluminescence (t-PL) decay kinetics are obtained for increasing pump fluences. Faster kinetics appear at higher pump fluences. These data are fit to some tri-exponential corresponding to X, XX, and XXX. But it is impossible to accurately obtain the desired amplitudes and time constants with this simple measurement that has no spectral resolution, and a time resolution of 100 ps. In our t-PL works, we employ a streak camera detection system with an optical trigger that enables spectral resolution as well as improvement in time resolution to 3 ps. This improvement from 100 to 3 ps is essential since the MX can decay on the 10 ps timescale. These more precise measurements also reveal a universal scaling curve, albeit with different slopes. In the case of CsPbBr3 CQD, we show data from different synthetic procedures, **Fig2c**, showing that synthesis can have as large an effect upon Auger lifetimes as CQD size.

**Fig2d** shows the relationship between the ASE threshold and volume, also adapted from Klimov[3]. indeed, there is excellent confirmation of the Standard Model that the gain thresholds scale with Auger recombination rate constants. But the universal curves are material specific, with CdSe and CsPbBr$_3$ CQD showing different responses. It would seem that the Standard Model is doing very well in terms of predicting outcomes. Further below we discuss the what it fails to reproduce and what is needed for a physically rigorous model that explains all phenomenology.

## 1.4. The Standard Model Provided the Design Principles for CQD Lasers

The impact of the Standard Model is that it created the design principles for CQD laser development. Standard laser development begins with assigning the system to being a two-level system (2LS) or a 3LS or a 4LS[56]. The Standard Model assumes that the CQD is a 2LS, with optical gain thresholds by SE requiring biexciton formation (N = 2). Given short Auger limited XX recombination times in CQD of ~ 30 ps, the aim of the CQD materials community was to slow Auger recombination by developing new core/shell structures. The Standard Model of CQD gain has led the entire CQD community down a path of slowing XX recombination as the main path towards efficient CQD lasers[2-4, 6, 12, 13, 18, 20, 57-74].

## 2. The Phenomenology of Optical Gain in Quantum Dots

### 2.1 The first step towards a laser: Measuring optical gain via stimulated emission

In the development of a CQD laser, the first step is to create optical gain via SE. The second step is to create ASE to create macroscopic gain in the medium. The final step is to place the gain medium in a resonator to form a working laser. To produce a clear path towards a Simple Model for optical gain that establishes well-founded design principles, it is helpful to begin with measurements of SE.

**Fig3.** presents an overview of the measurement of SE. All other measurements of SE employ pumping into the excitonic continuum at 400 nm for the sake of experimental convenience[3, 6, 11, 17, 18, 57-59, 61, 62, 64, 69, 72, 74-85]. Instead, we have employed state-resolved optical pumping using tunable pump pulses[86-88]. The idea is shown in **Fig3a**. Shown is a linear absorption spectrum and PL spectrum of CdSe CQD in dispersion. The CQD may be pumped with tunable pulses that are resonant with specific excitonic states as shown by the arrows. Also shown is the non-linear absorption spectrum ($OD_{NL} = OD_0 + \Delta OD$). The non-linear absorption spectrum corresponds to the spectrum of the pumped sample. The absorptive features are bleached nearly to the point of transparency, illustrating the filling of the band edge exciton level. At the red edge of the spectrum, it becomes negative. This negative absorption corresponds to optical gain via SE. Upon inspection two points become clear: 1) the SE exists at the red edge of the spectra, 2) the SE cross section is small relative to the linear absorption cross section.

**Fig3b**. shows the SE spectra in more detail, employing state-resolved optical pumping. Based upon the initially pumped state there are two clear observations: 1) the cross section depends on the initially pumped state, 2) the SE gain bandwidth depends on the initially pumped state. With simple 400 nm pumping as done in all other SE experiments, these effects would not be possible to observe. These two observations suggest that the physics of optical gain is much more complex than the Standard Model has one assume.

To further illustrate how state-resolved optical pumping reveals key new observations, the gain thresholds and cross sections are shown in **Fig3c**. Shown are data for CdSe and CdSe/ZnS CQD. **Fig3c** (top) quantifies the gain threshold vs initially pumped state. It is clear that the gain threshold increases for higher excess electronic energy, and that shelling with ZnS lowers thresholds across the spectrum. **Fig3c** (bottom) quantifies the SE cross section vs initially pumped state for the two CQD. **Fig3d** shows the CQD size dependence of the SE threshold, upon pumping into the band edge exciton. Using state-resolved optical pumping, we recover the true gain performance of the CQD, with nearly size independent gain thresholds. Closer inspection reveals that the gain threshold slightly increases, moving from smaller bluer CQD to larger redder CQD. This response we observed is in stark contrast to all other works employing 3.1 eV excitation. Those works all reveal that there is a strong size dependence to the SE threshold with smaller CQD having higher thresholds, and in many cases supporting no gain at all. As a result of the believed poor performance of smaller CdSe CQD, efforts were made to develop other compositions of CQD that could support SE in the blue.

In the Standard Model there are only two states or levels. Moreover, there are no trap states which are known to exist in real CQD[47, 89-98]. We proposed that the threshold increased with excess electronic energy due to hot exciton effects. Specifically, we proposed that hot excitons can undergo efficient trapping to surface states that compete with hot exciton cooling to the band edge exciton. The ideas of hot exciton surface trapping has been discussed in detail, and can be measured by TA measurements[47, 93, 99, 100] as well as PL excitation (PLE) spectra[93, 94, 98, 99]. These works quantify hot exciton trapping to surface states. Upon population of surface states, there are new excited state absorptions arising which block the desired SE[89, 100]. In this manner, the state-into which one pumps bears critical impact upon the optical gain performance of real CQD. This initial state dependence is not present in the Standard Model.

The SE performance in response to state-resolved optical pumping is most dramatically shown in graded alloy CdSe/CdS/ZnS CQD[95-97]. The SE response is shown in **Fig3e-f,** with pumping into $X_1$ and $X_3$ states. The response to pumping into $X_1$ is qualitatively similar to the response of standard CdSe cores. But upon $X_3$ pumping, the differences are striking. Pumping into this higher exictonic state enables formation of a statistical distribution of MX, past the XX state. The graded alloy CQD notably enables SE from higher excitonic states, to the blue of the spontaneous PL or band edge exciton absorption.

Using state-resolved optical pumping, it is clear that band edge pumping is the best for finding the lowest thresholds and largest SE cross sections. With this pumping scheme, we generalized the gain threshold for CQD from red to green to even blue, using CdSe cores in **Fig3d**. State-resolved optical pumping yields a size independent gain threshold in stark contrast to non-resonant pumping into the excitonic continuum at 400 nm. With 400 nm pumping, the gain thresholds are strongly size dependent[11, 18, 76], with smaller CQD showing no development of SE due to mysterious parasitic loss mechanisms that we have since explained as arising from hot exciton surface trapping[47, 88, 94, 98, 99].

## 2.2. The Second Step Towards a Laser. Amplified Spontaneous Emission Measurements

There are two ways in which optical gain is measured. Stimulated Emission (SE) measurements are primary measurements in that they reveal the fundamental parameters that govern optical gain performance: threshold, cross section, bandwidth, lifetime. A secondary measurement of optical gain is Amplified Spontaneous Emission (ASE). ASE is a time-integrated measurement of optical gain performance, driven by pulsed lasers. Since there is no probe pulse to measure the buildup of SE, there is only the buildup of a spontaneous emission event via ASE. ASE is the most commonly used method of characterizing optical gain thresholds and cross sections due to its simplicity of implementation.

The observation of ASE is critical to laser development as ASE is one of the two main processes required to make a laser. The final step being placement of the gain medium in an optical cavity to form an oscillator. Due to the primacy of the ASE measurement as well as due to its relative experimental simplicity, it is the most common means to quantify optical gain performance. Hence a critical analysis of ASE measurements is essential towards understanding their role in informing CQD laser development. We provide a brief overview of ASE here, with further discussion in the modeling section below.

**Fig4** shows ASE spectra of different CQD to illustrate the nature of the signals. **Fig4a** shows spectra from lead-halide perovskite (LHP) CQD[69]. In each case, the ASE gain spike appears to the red of the PL from the band edge exciton. This observation of redshifted ASE is taken as assignment that the light amplification is taking place from biexcitons not excitons[2, 3, 12, 13]. A key observation is that In nearly all ASE measurements, the gain spike appears on the red edge of the spontaneous PL bandwidth.

**Fig4b** shows ASE data on an inverted core/shell CQD by Klimov[101]. The objective of this CQD system is to create a biexciton binding energy that is negative rather than positive, as is the norm. This question of biexciton formation and binding energies will be discussed in detail below. The main point of this ASE experiment on an inverted core/shell structure is that the ASE gain spike appears at the center of the spontaneous PL band, consistent with ASE is taking place due to X rather than XX, which is a significant step forwards towards optical gain with low thresholds and long lifetimes. While such metrics were achieved with this inverted core/shell structure, it also has disadvantages due to weak oscillator strength of a spatially separated exciton.

State-resolved optical pumping can also be done for ASE measurements and remains similarly valuable to inspecting gain phenomena and understanding the underlying physics. **Fig4c** shows the ASE from CdSe CQD with 1S and 1P pumping[87, 88]. The first observation is that CdSe CQD indeed show ASE at the center of the PL band, showing gain from single excitons rather than biexcitons. The ASE with X1 pumping even appears at the blue side of the PL, which is unanticipated. The most significant observation is that the position of the ASE gain spike in a steady-state experiment is sensitive to MX formation. The presence of MX controls the SE spectra and even the ASE spectra. This gain bandwidth control may be used for all-optical modulators and logic[102]. These ASE effects cannot be explained by the Standard Model.

The spectral position of the ASE spike will be even more revealing as we discuss later. For the purposes of quantification of gain performance by ASE, the metric is the threshold. The total amplitude is rarely reported, however. **Fig4d-f** show the ASE threshold response for different conditions. **Fig4d** shows the response for CdSe CQD[20]. There is a strong size dependence to the ASE response in contrast to TA measurements which reveal that the SE performance is size independent[87, 88]. The ASE response shows that the threshold and slope have a strong size dependence, with larger CQD showing better performance. This preference for red CQD ASE in CdSe systems has led the effort to produce new CQD systems towards lasers in the blue[18]. As we discuss further below with our Simple Model, the gain threshold for an ideal CQD should be size independent, for both SE and ASE experiments.

**Fig4e** shows the ASE response for the core/shell CQD. There is a characteristic initial rise of the PL, followed by a faster rise in the ASE. This core/shell CQD shows SE at <N> = 0.7, but the ASE threshold is 2 mJ / $cm^2$, which is not especially small. This divergence further suggests potential conceptual problems with the Standard Model.

**Fig4f** shows the ASE response for CdSe, pumped into the 1S and 1P states, using state-resolved optical pumping[87, 88]. Again, the ASE thresholds are not particularly low, even with SE threshold of <N> = 1.3. A simple but significant aspect of ASE measurements is that they are subject to the quality of the film which induces losses that block the desired gain. There is no such problem for SE measurements, hence one potential difference is solved. But there are more fundamental spectroscopic reasons for the differences as discussed below. Focusing on the relative difference based upon initial excitonic state, higher states have higher ASE thresholds and redder ASE spectra.

## 2.3. Measuring True Gain Thresholds via Stimulated Emission: Exciton or Biexciton Gain

While ASE measurements can produce thresholds in terms of fluence, they cannot identify the exact gain threshold in terms of number of photons absorbed, and excitons created. A more precise measurement of gain thresholds is obtained via SE rather than ASE measurements. The SE measurements reveal the intrinsic gain performance of the material. Whereas ASE measurements reveal the gain performance convoluted with losses due to film quality.

The first measurement of gain thresholds in colloidal CQD were in CdSe CQD by Klimov[11]. Subsequent efforts to verify the universality of gain and any size dependence of gain performance have shown that gain thresholds are strongly size dependent[76]. Based upon non-resonant pumping at 400 nm, it was observed that smaller CQD had larger gain thresholds. The smallest CQD in the blue were not able to support optical gain at all. This size dependence presented a "blue wall" for CQD gain thereby creating the need for new materials solutions[14, 18, 71, 84].

Using state-resolved optical pumping with $X_1$ excitation, we examined the gain thresholds of different sizes of CdSe, as well as the influence of shelling via hard interfaces of ZnS and graded alloy interfaces of ZnS, **Fig5a**. Using $X_1$ pumping, the gain threshold, <N>, is < 1.5 for CdSe and CdSe/ZnS. This gain threshold is totally inconsistent with the proposal of a biexciton being a threshold for gain with N = 2. The observed gain SE threshold suggests that the CQD may be a 3LS, consistent with the presence of a PL Stokes shift. To lower the gain threshold to <N> < 1, there have been a variety of strategies implemented. **Fig5b** shows the SE threshold for the inverted core/shell CQD with negative biexciton binding energy, and **Fig5c** sows the curves with the effects of electrochemical charging included. In both cases, the SE thresholds are reduced from the Standard Model anticipated threshold of N = 2, to smaller numbers of <N> < 1.

These data represent measurements of the intrinsic SE threshold. But they are reported in two different ways. One can measure the buildup of SE as $OD_{NL}(E, t; <N>) = OD_0 + \Delta OD(E, t; <N>)$, in which the negative part of the nonlinear absorption spectrum is the SE spectrum. One can also report the fractional bleaching $\Delta OD(E, t; <N>) / OD_0$. The former method is more precise because it does not involve division by a small and noisy denominator. And it is more informative in that it provides the undistorted SE spectrum, and its amplitude, both of which enable modeling of the ASE response. The main result for gain characterization is that SE is a more precise measurement of threshold than ASE, which we will discuss in greater detail below.

## 3. A Simple Model Explains and Predicts Optical Gain Processes in Quantum Dots

## 3.1. Is a CQD a Two, Three, or Four Level System: Predicting Outcomes

Having observed the linear spectroscopy and non-linear optical gain performance of CQD, a rigorous model begins with classification of the CQD into a two-level system (2LS), or three-level system (3LS), or a four-level system (4LS) or higher[56]. The presence of an excitonic Stokes shift guarantees that the CQD cannot be a 2LS. Hence the very starting assumption of the Standard Model is incorrect. The Stokes shift guarantees that a CQD must be at least a 3LS as a lasing scheme.

The relevant states and transitions of these lasing systems are shown in **Fig6a**. The level diagrams are shown with inclusion of a configuration coordinate representing lattice motions. In atomic systems the configuration coordinate is not present, and one has a true level structure. But in the case of a CQD, one must consider the presence of coupling to lattice phonons. In this picture, the 2LS consists of an occupied ground state and an unoccupied excited state, the exciton level. A 2LS is formed by having zero relative displacements between the excited and ground state potential energy surfaces, resulting in zero coupling to phonons. In this 2LS, the absorbing and emitting transitions are at the same energy.

Since there is a PL Stokes shift between the absorbing and emitting states, the CQD must at least be a 3LS. Now the emissive transition is redshifted with respect to the absorptive transitions. There are two excited states, $X_A$ and $X_E$. Here, the upper exciton level is split due to known fine structure effects [1, 43-45, 103], giving rise to three levels. Here there are transitions from 1 → 2 → 3 → 1. Upon introduction of exciton-phonon coupling by displacing the excited state potential horizontally, the simple 2LS now becomes a 4LS. Here there are transitions from 1 → 2

→ 3 → 4 → 1. The 4LS has phonon progressions giving rise to new levels, but not from distinct electronic states. Now if the 3LS of a real CQD has inclusion of exciton-phonon coupling, it will produce a 5LS.

In order to model the CQD response these level systems need to be characterized in terms of the splittings and transition rate constants. The splittings arise from the Stokes shifts. And the transition rates constants can be obtained from time-resolved measurements. **Fig6b** shows t-PL transients for CdSe CQD excited at 400 nm. The transients are shown at select probe photon energies. The two energies to focus upon are 2.3 – 3.0 eV, and 1.8 to 2.0 eV. The higher energy PL band corresponds to hot PL from relaxing excitons. The lower energy PL band corresponds to emission from the X1 band edge exciton. The data obtained with a 3 ps instrument response function (IRF) shows that on the 1 -2 ps timescale there are still unrelaxed hot excitons. The appearance of the X1 band edge PL takes time to appear, also on the timescale of 1 – 2 ps. Hence we can estimate that the timescale for the 400 nm pumped state to relax to the emitting state is 1 – 2 ps at most.

Another measurement of the relaxation timescale is SE transients, **Fig6c**. Shown here are SE transients for CdSe CQD with $X_1$ pumping directly into the band edge exciton. Here there is no relaxation from the 400 nm non-resonant continuum to the $X_A$ state as the pump directly populates that state. The data shows the SE has a buildup time of 500 fs. Hence the timescale of evolution from $X_A$ to $X_E$ is now known. From the PL spectroscopy, as well as t-PL and TA spectroscopy, one must conclude that the colloidal CQD must be at least a 3LS and based upon exciton-phonon coupling, may form a 5LS.

The gain kinetics of the 2 / 3 / 4LS are shown in **Fig6d**. This simulation corresponds to continuous-wave (cw) pumping which is helpful for low power laser development. As expected from standard laser kinetics[56], the 2LS will never produce gain with cw pumping. The 3LS will produce gain with a finite threshold. And the 4LS is the ideal lasing system in that it supports thresholdless gain.

## 3.2. Adding Biexcitons to Complete the Level Structure

According to the Standard Model, a biexciton (XX) is required to produce optical gain in CQD by virtue of their being a 2LS. Since the real CQD is not a 2LS, a biexciton is not necessarily relevant to the production of optical gain via SE. Instead, the biexciton creates an absorptive signal that creates loss instead of gain. Hence the biexciton should be discussed, both as a gateway and a gatekeeper for generating optical gain.

The essential ideas of biexcitons in CQD are presented in **Fig7**. In particular, **Fig7a** shows the salient energy level diagrams needed to rigorously evaluate how XX can either block or allow gain. In the simplest case of the CQD as a 2LS, the idea is that excited state absorption (ESA) from X → XX is resonant with the desired SE and thus blocks SE from developing. Since the SE has the same transition energy as the linear absorption into X, the nonlinear absorption into XX is also at the same energy. In a 3LS, the SE is redshifted with respect to the linear absorption into X. To block the desired SE, there should be ESA into XX at this same energy. With these energetics, XX is characterized by a positive binding energy, $\Delta_{XX} = E_{2X} - E_{XX}$. Alternatively, the binding energy may be defined as $\Delta_{XX} = E_{0,X} - E_{X,XX}$. In the 4+LS, the same idea holds as for the 3LS in that the ESA into XX blocks the desired SE from X.

The gain kinetics under steady-state conditions with inclusion of ESA into XX are shown in **Fig7b**. The main result is that even the 3LS and 4LS do not produce gain due to perfect cancellation of SE via ESA. Hence biexcitons are first and foremost a gatekeeper to efficient optical gain ad the single exciton level. Since biexcitons are fundamental to CQD, they cannot be engineered "out". Instead, the ideal biexciton interactions to enable efficient gain, should be developed.

The formation of biexcitons in CQD has been well reviewed[30, 104, 105], based upon observation by t-PL[52, 53, 55], TA[46-48, 50, 52, 100], and Coherent Multi-Dimensional Spectroscopy (CMDS)[51-53, 106]. The simplest idea of XX in CQD was that they were bound by some binding energy. Then it was observed that the binding energies obtained via emissive experiments (t-PL) were not identical to the binding energies obtained via absorptive experiments (TA and CMDS). This disconnect gave rise to our proposal of a fine structure to XX and a Stokes shift for XX, precisely as known to exist for X[48]. Furthermore, using state-resolved optical pumping, we have shown that there is a coarse as well as a fine structure to XX[50]. Hence the excitonic structure of biexcitons are real and as rich as for excitons. Yet this structure to XX is rarely discussed, with XX being observed largely by simple t-PL measurements.

Representative data on biexciton formation in CQD are shown in **Fig7c** for TA experiments. The TA data is from CdSe CQD pumped directly into the $X_3$ state consisting of a 1P exciton. A TA measurement of XX binding energies is more subtle than the simpler t-PL measurements. In a TA experiment, the change in absorbance signal arises from three fundamental processes: ground state bleaching (GSB), stimulated emission (SE), and excited state absorption (ESA). The GSB and SE signals arise from X and the ESA signal arises from XX. Hence an experimental photo-induced absorption signal arises from ESA from X → XX.

Such an induced absorption signal is revealed in the TA spectra to the red and blue of the band edge exciton bleaching signal[38, 46, 47, 50, 52, 99]. The induced absorption signal rises on the 200 fs timescale and decays on the 1 ps timescale. The rise and fall of this signal reflects the process of biexcitons cascading down their level structure towards the band edge biexciton. The main point is that this ESA into XX creates a loss process which can block the desired SE and optical gain[30]. For CQD that are not well passivated or are excited into the continuum at 400 nm, there will be hot exciton surface trapping as well[30, 47, 89, 93, 94, 98-100]. This trapping to surface excitonic states gives rise to additional PIA loss mechanisms which block the desired gain.

Representative data on biexciton formation in $CsPbBr_3$ CQD are shown in **Fig7d** for t-PL experiments [54, 55]. The use of t-PL to monitor the formation of multiexcitons and their interaction energies and kinetics of recombination appears ideal in that it is the simplest experiment and the easiest to analyze and interpret. The problem with t-PL measurements is that they have a worse time resolution, with an instrument response function (IRF) of 3 ps at best, and closer to 100 ps for most measurements. Hence t-PL cannot observe the structure of MX, which can be observed in absorptive experiments like TA and CMDS.

Data is shown from $CsPbBr_3$ CQD at an early time delay of 5 ps. Shown are four emission bands arising from X ($1S^1$ → 0), XX ($1S^2$ → $1S^1$), MX lower ($1S^21P^1$ → $1S^11P^1$), and MX upper ($1S^21P^1$ → $1S^2$). Clearly the spectra overlap, and there is no way to fit the experimental spectrum to four peaks without ambiguity. This ambiguity problem is solved by performing a global fit of all data over a wide range of pump fluences and time delays. With this global fitting procedure, we can observe MX and their kinetics in t-PL experiments[54, 55]. In this case the MX are emissive states rather than absorptive states. Hence MX formation gives rise to emissions, especially SE, at new energies. Such emission from MX gives rise to gain bandwidth control in SE experiments.

## 3.3. Modeling Stimulated Emission Thresholds and Cross Sections: Rationalizing Existing Results and Predicting New Phenomena

Upon introduction of ESA into XX as well as SE from XX, the onset of SE can now be modeled to rationalize existing observations and to predict new phenomena, **Fig8**. Model spectra are shown in **Fig8a**. For the sake of simplicity, we model the lineshapes as a Gaussian, representing all homogeneous forms of line broadening from excitonic fine structure, to phonon progressions. Sample heterogeneity is not included. Shown are the linear absorption from $|0\rangle \rightarrow |X_A\rangle$, nonlinear absorption $|X_A\rangle \rightarrow |XX_{A1}\rangle$, emission from $|X_E\rangle, \rightarrow |0\rangle$, nonlinear absorption $|X_E\rangle \rightarrow |XX_{A2}\rangle$, and emission from $|XX_E\rangle, \rightarrow |X_E\rangle$.

The linear absorption, and PL from X and XX are both directly measured quantities, $A_X$, $PL_X$, $PL_{XX}$, respectively. In contrast the non-linear absorption must be inferred by summing the three terms in a TA signal: ground state bleaching, stimulated emission, excited state absorption into XX[30, 47, 48, 50]. The nonlinear absorption can be indirectly obtained prior to relaxation from the absorbing state to the emitting state, $A_{XX1}$, or after relaxation, $A_{XX2}$. A useful measure of the nature of the biexcitons formed is the biexciton binding energy. The binding energy is taken as the energy of two excitons minus the energy of a biexciton, $\Delta_{XX} = 2A_X - A_{XX}$. The emissive biexciton binding energy can be directly measured, $\Delta_{XX}^E = PL_X - PL_{XX}$. The absorptive binding energies are computed based upon the relationship between the three signal terms, $\Delta OD = GSB + SE + ESA$.

For pulsed excitation, we treat the excitation process in the Rabi picture. Here, there is a ground state and a doubly degenerate excited state. Because excitation is not into the continuum at 400 nm, it is not appropriate to assume a Poisson distribution as is the norm in the Standard Model. The Standard Model invokes the mean exciton occupancy is equal to the produce of the absorption cross section and pump fluence, $<N> = \sigma J$. The Rabi model reproduces the experimental saturation curves obtained by both TA[106] and t-PL[54] with band edge excitation, **Fig8b**. Data is shown for $CsPbBr_3$ CQD. Based upon the physical suitability of the Rabi model, and its excellent reproduction of the experimental data with band edge pumping, we can use this model to more precisely simulate the excitation intensity dependence of the gain response.

**Fig8c-d** show the simulated SE response as a function of mean exciton density, $<N>$, for the 3LS and 4LS. The CQD gain material is modeled as a thin film < 0.1 mm, so propagation effects and MX recombination kinetics are not relevant since the transit time is < 300 fs. The model includes loss due to a pumped CQD sample. The 2LS is omitted since it does not reproduce the most basic aspect of the linear spectroscopy: the Stokes shift. The first relevant model of CQD gain is the 3LS without ESA into XX. The contour plot shows the SE spectra as a function of mean excitation number, $\sigma_{SE}(E, <N>)$. The top panel shows the projection of the SE spectra for $<N> = 1$ and 2. The panel to the right shows the projection of the excitation spectra, obtained at the energy of the SE threshold and the SE maximum.

The 3LS shows a small single exciton gain spectrum, at the red edge of the PL band due to overlapping absorption and emission spectra as is the norm for all CQD. The 3LS shows a large maximum gain of 1, relative to the absorption cross section. The 4LS shows a similar response, but with more single exciton gain amplitude, and a threshold approaching zero, as is the theoretical limit for a 4LS[56].

The problem is that this model is not realistic since all CQD have ESA into XX, rather than having two uncorrelated excitons. Given that XX and MX are fundamental excitations in a CQD, they must be introduced on equal footing in any realistic model. The idea of involving XX is that there is both ESA from X → XX as well as SE from XX → X. Based upon the transient absorptive[30, 46, 47, 50-52, 100, 106] and emissive[52, 54, 55] measurements of biexcitons, we propose there are distinct binding energies for both transitions. The distinct binding energies are made possible by a manifold of biexciton states as a consequence to the manifold of single exciton basis states. The SE from X will be partially cancelled by ESA into XX. And what will remain is SE from XX, with minor contributions from X. In short, a 3+LS with biexciton formation is the appropriate minimal model of the CQD as a lasing system. Indeed, in 2009 we proposed that the CdSe CQD is better modeled as a 3LS[87, 88].

**Fig8e-f** show the gain response for the two salient lasing systems, with addition of ESA into and emission from XX. The effect of introducing ESA into XX as a loss mechanism is clearly seen in the differences between the top row (no ESA) and the bottom row (with ESA). This loss channel raises the SE threshold from <N> = 1.0 to 1.5 for the 3LS, and from 0.5 to 1.0 for the 4LS. A predicted threshold of <N> = 1.5 is precisely what we observed with band edge pumping in red, green, and blue CdSe CQD [87, 88]. We observed a threshold of <N> = 1.0 with graded alloy CdSe/ZnS

CQD [95-97], suggesting that biexcitonic loss effects are weaker in this system. The 3LS with ESA into XX quantitatively reproduces the SE thresholds for CdSe based CQD. Further below we discuss the possibility of transforming the CQD from a 3LS to a 4LS, further lowering the gain threshold.

**Fig8g-h** shows how the SE thresholds are controlled by the biexciton binding energies relative to the Stokes shifts. Shown are the response of the 3LS and 4LS vs the absorptive biexciton which causes loss, $A_{XX2}$, with a binding energy of $\Delta_{XXA2}$. The responses are computed at the energy of the SE threshold and the SE maximum, which are not necessarily identical even for these simple Gaussian spectra.

The gain threshold by SE measurements is the primary metric of CQD performance for optical gain. An often overlooked aspect of CQD gain performance is the cross section. As discussed above in the phenomenology section, the SE cross sections are very small. There are two ways to measure the SE cross section. One can measure the non-linear absorption spectrum, $OD_{NL}$, and monitor the <N> at which the spectrum begins to show negative absorption meaning optical gain. One can also measure the fractional bleaching at the position of the SE, $\Delta OD/OD_0$. The latter method is commonly used because it exaggerates the effect of the onset of SE.

The problem with that method is that a large numerator is being divided by a small and noisy denominator resulting in poor quality spectra that accentuates measurement errors. This $\Delta OD/OD_0$ spectrum moreover does not directly show the desired SE spectrum. We propose that it is a better method to plot the $OD_{NL}$ spectrum because it removes the noisy denominator problem, and more importantly it shows the exact cross section and bandwidth which are the key metrics for laser performance.

Now examining the SE cross sections from simulations, it is clear that loss by absorption into XX attenuates the magnitude by a large amount. The predicted SE cross sections in the absence of ESA are $|\sigma_{SE} / \sigma_A| = 1$, precisely as observed for atomic and molecular lasing systems that do not support biexciton formation[56]. Upon introduction of biexciton formation, the SE cross section drops to 0.3. The vast majority of SE measurements of CQD gain obtain a ratio of 0.05 - 0.10, with strong size dependence. We observed a ratio of 0.10 in CdSe CQD for all sizes[87, 88], showing the value of band edge pumping which bypasses hot exciton surface trapping. In graded alloy CdSe/ZnS CQD[95-97] we obtained a cross section of 0.18, and also in core/barrier/shell CdSe/ZnS/CdSe CQD[107] we obtained a cross section of 0.13. For the graded alloy CQD, the SE cross section became as large as 0.36 with pumping into higher excitonic states. Introduction of loss from XX well reproduces the presence of small SE cross sections that can be increased based upon shelling.

## 3.4. The Promises and Pitfalls of Amplified Spontaneous Emission Measurements

Whereas SE measurements of optical gain are the most precise and information rich methods, the far more common method of characterizing optical gain performance is steady-state ASE measurements. The ASE experiment is more commonly used because it is far simpler, employing only a single pump beam. The ASE measurement is also important towards demonstrating the suitability of CQD for lasers since ASE the first of two main processes to make a laser. The second process is to create optical feedback by placing the gain material in a resonator to form an oscillator.

The simulations of SE spectra for model systems successfully reproduces experimental thresholds and cross sections. From these SE spectra, the ASE spectra can also be simulated as shown in **Fig9**. The ASE spectra are obtained directly from the SE spectra. It is assumed that there are no material losses due to film quality.

The response of the 3LS with and without ESA into XX is shown in **Fig9a-b**. The ideal 3LS The first observation is that the ASE threshold increases from 0.27 to 0.69 mJ / cm$^2$, approximately a factor of two difference. Of course, the ASE threshold with ESA into XX will be dependent upon the relevant XX binding energy. The second observation is that the ASE spike appears at the center if the PL$_X$ spectrum for the 3LS but is redshifted upon inclusion of ESA into XX. The appearance of the ASE spike at the center of the PL band is consistent with the results on CdSe using state-resolved optical pumping into X$_1$, and the appearance of the ASE spike on the red edge of the PL band is consistent with all other experiments done with pumping at 3.1 eV which creates losses that redshift the ASE spike. A third observation is that the total amplitude of the ASE spike depends on the loss due to XX, with a factor of two difference. The fourth and final qualitative observation is the spectral shifting of the ASE spike with increasing pump fluence. This effect arises due to the interplay between gain and loss of excited vs unexcited CQD.

The 4LS is shown in **Fig9c-d**. The 4LS without ESA is similar to the 3LS but the threshold drops to 0.06 mJ / cm$^2$, and with ESA is 0.3 mJ / cm$^2$. The 4LS shows a richer spectral response, with a strong blueshift with fluence for the pure 4LS, and a strong redshift with fluence for the 4LS with ESA into XX.

**Fig9e-f** shows the predicted ASE response of the CdSe/CdS/ZnS graded alloy CQD for which the SE spectra are shown in **Fig3e-f.** The predicted response of the real CQD is remarkably consistent with the 3LS with ESA into weakly bound biexcitons. Both show the identical gain threshold of 0.27 mJ / cm$^2$. Both show a blueshifting of the ASE spike with increasing pump fluence. And both show the ASE spike at peak gain appearing at the center of the PL band. The ASE response upon $X_3$ pumping is strikingly different. There are now two main ASE spikes, with a continuum between the two. The first ASE spike has a wider bandwidth than with $X_1$ pumping. And the second ASE spike does not exist with $X_1$ pumping.

These simulations reveal that the ASE threshold and spectral response are complex functions of the excitonic and biexcitonic spectral positions, with absolutely no dependence on XX Auger lifetime. So it remains to be addressed why the Standard Model found a relationship between ASE threshold and CQD volume. As shown in **Fig2**, the ASE threshold correlates with CQD volume thereby creating a semi-universal curve. It is semi-universal in that there are different families of curves for different materials, such as CdSe vs CsPbBr$_3$ CQD.

**Fig10a** shows the semi-universal curves relating absorption cross section to volume. The curves for CdSe were taken from [108], and the data for CsPbBr3 were taken from [109]. Since these data were not obtained by the same methods, a more quantitative analysis requires simultaneous measurements on both classes of CQD using the same methods. We did so for one size od each CQD [55], with points shown as well. The CdSe curve and the CsPbBr3 curve were then overlaid with the single points that were simultaneously obtained. Scaled in this manner, there are now quantitatively accurate curves for both families of CQD. The LHP CQD have a larger absorption cross section per unit cell, which gives rise to their many remarkable optical properties [110-114].

**Fig10b** shows the optical gain thresholds from both Stimulated Emission (SE) and ASE measurements for a variety of sizes and compositions of CQD. The threshold is reported as $<N>_{threshold}$ rather than $J_{threshold}$ to better show universal behavior. We begin the observations focusing on SE measurements since they are more accurate than ASE. Our work[87, 88] using band edge excitation on CdSe CQD shows a size independent $<N>_{threshold} = 1.5$, consistent with our modeling of a 3LS with ESA into XX that is strongly bound. In contrast, SE from Klimov using 3.1 eV pumping[76] shows a strong size dependence, with bluer dots not supporting SE. This strong size dependence has resulted in a "blue wall" for CQD lasers[115], purely based upon the problems of 3.1 eV pumping. The ASE data from Klimov[2, 3] is more scattered due to the imprecision of the measurement. But the ASE threshold for CdSe CQD is ~ 1.5, consistent with the SE data from our group. Our SE measurements on CdSe/CdS/ZnS graded alloy CQD[95-97] reveals a threshold of 1, illustrating a gain threshold reduction due to biexciton effects. Changing composition from CdSe to $CsPbBr_3$, the ASE threshold is lowered to 1, in data from both Klimov[2, 3] and Kovalenko[69]. With SE and ASE thresholds plotted in this manner, it is clear that there is universal size independent behavior, and the more important result that multi-exciton Auger recombination has absolutely no impact upon optical gain thresholds.

Also shown in **Fig10b** are additional curves that are predictions from the simulations. The Universal ASE threshold curve can be lowered by a factor of 3x by engineering the XX interactions suitably. The Universal ASE curve can also be lowered by 10x by using light harvesting shells [82, 107, 116] to increase the absorption cross section for excitation, without changing the emissive properties. The ultimate goal is to realize thresholdless gain in CQD which may be approached by transforming the 3LS of current CQD to a 4LS or 5LS which would achieve this longstanding goal, as shown by the lowest curve of gain threshold.

The results of **Fig10** provide an understanding of the past 25 years of CQD gain development, and what we can look forward to in the future. Since XX Auger lifetimes have no influence upon optical gain performance, the entirety of the past 25 years of work has been process control towards better quality films. Improvements in CQD syntheses have resulted in better control of hot exciton trapping which result in optimal SE performance, and in CQD film fabrication to result in optical ASE performance – given the fundamental constraint of the absorption cross section. The only way past this wall is to employ the strategies we propose here.

## 4. Summary and outlook

Since optical gain in CdSe CCQD was first reported in 2000, the field has exploded. A Standard Model for optical gain in CCQD was developed based upon phenomenology. This Standard Model predicts that the XX Auger recombination timescale is the primary measure that predicts ASE thresholds. The literature on ASE appeared to validate the Standard Model. But the Standard Model cannot reproduce any of the SE behavior of CCQD. A Simple Model was proposed based upon the known exciton and biexciton state energies. In this Simple Model which follows the response of 3 and 4 LS dressed with XX, all SE and ASE phenomena are now explained, including the physical origin of the gain thresholds. The thresholds are determined by the energies of the transitions and not XX recombination timescale. The ASE threshold is shown to follow a universal curve as a function of absorption cross section. This universal curve represents the fundamental limit to the gain performance. The Simple Model predicts that there are new avenues to circumvent this universal curve by employing excitonic engineering of XX interaction energies and / or absorption cross section by light harvesting shells. Based upon the correct physical inputs to evaluating optical gain efficiency, a path forward is laid to quantum leaps in optical gain performance in CCQD.

## AUTHOR INFORMATION

**Corresponding Author**

* Email: pat.kambhampati@mcgill.ca, Phone: +1-514-398-7228

**Author Contributions**

The analysis and modeling was done by DZ. The writing was done by PK and DZ.

**Funding Sources**

PK acknowledges financial support from NSERC. DZ acknowledges fellowship from McGill University.

**Notes**

The Authors declare no competing financial interests.

**Figures.**

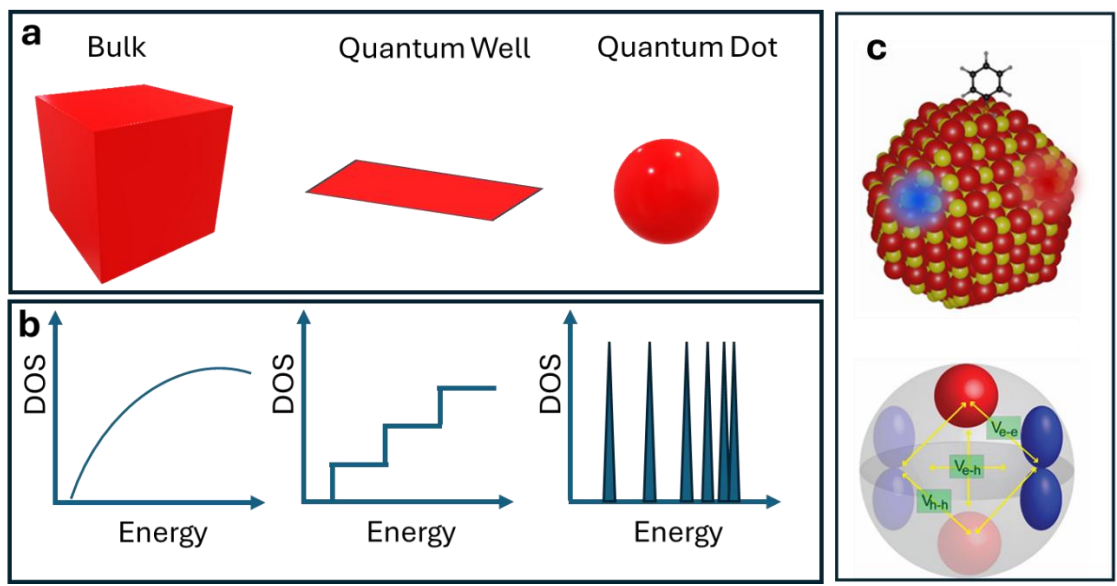

**Fig1. Schematic illustration of quantum confinement effects on electronic structure of semiconductor particles.** a) Illustration of three different confinement classes. b) Illustration of the electronic density of states for each class of confinement. c) Illustration of a CdSe CQD with atomistic detail (top), and illustration of one of many possible biexciton states (bottom).

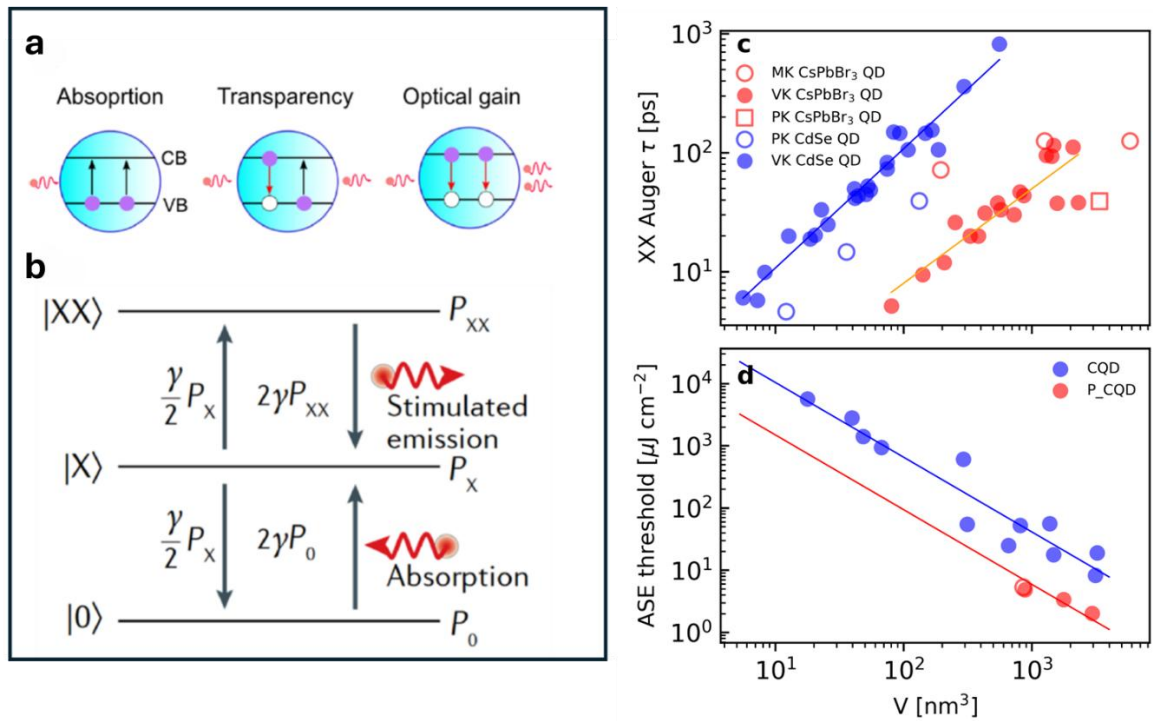

**Fig2. Illustration of the Standard Model of optical gain in CQD.** a) The Standard Model invokes a two-level system to describe the physics. b) Transitions from excitons to biexcitons have been written in terms of pumping and recombination rate constants. c) The relationship between the biexciton Auger lifetime and CQD volume. d) The relationship between Amplified Spontaneous Emission (ASE) threshold and CQD volume. Figures and data adapted from [3, 5]

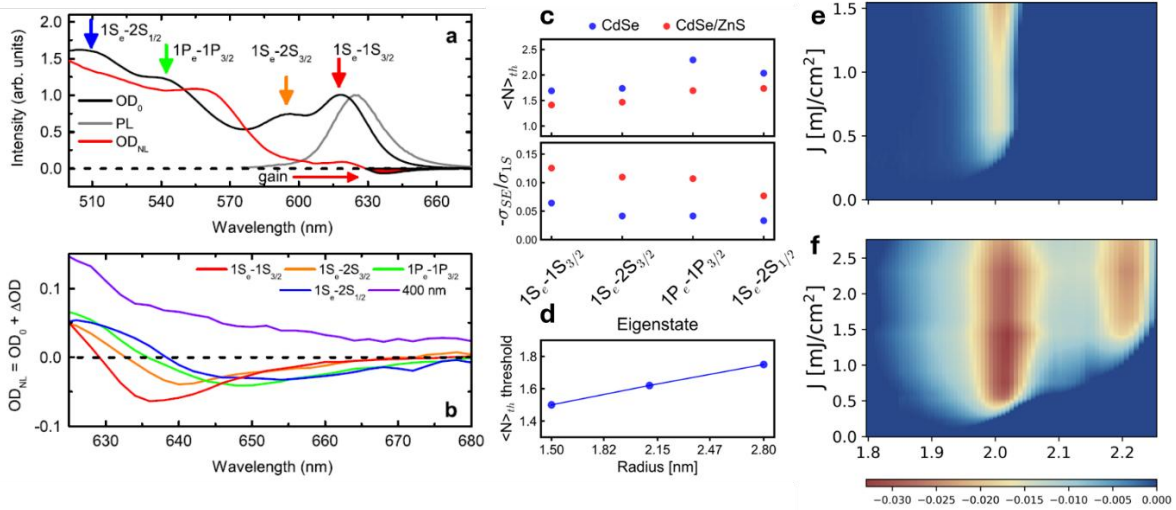

**Fig3. Measuring optical gain via excitonic state-resolved optical pumping.** a) Overview of Transient Absorption (TA) measurements of optical gain as revealed by the appearance of Stimulated Emission (SE) in CdSe CQD. Different pump wavelengths are used to better understand optical gain physics. Adapted from [87]. b) The SE spectra strongly depend upon the initially prepared excitonic state. Adapted from [87]. c) The gain thresholds obtained via state-resolved optical pumping (top) and gain cross sections (bottom). Adapted from [88]. d) The size dependent gain thresholds obtained via pumping into the 1S band edge exciton. Adapted from [88]. e) – f), The SE spectra of a CdSe/CdS/ZnS graded alloy CQD with $X_1$ and $X_3$ pumping, respectively. Adapted from [97]

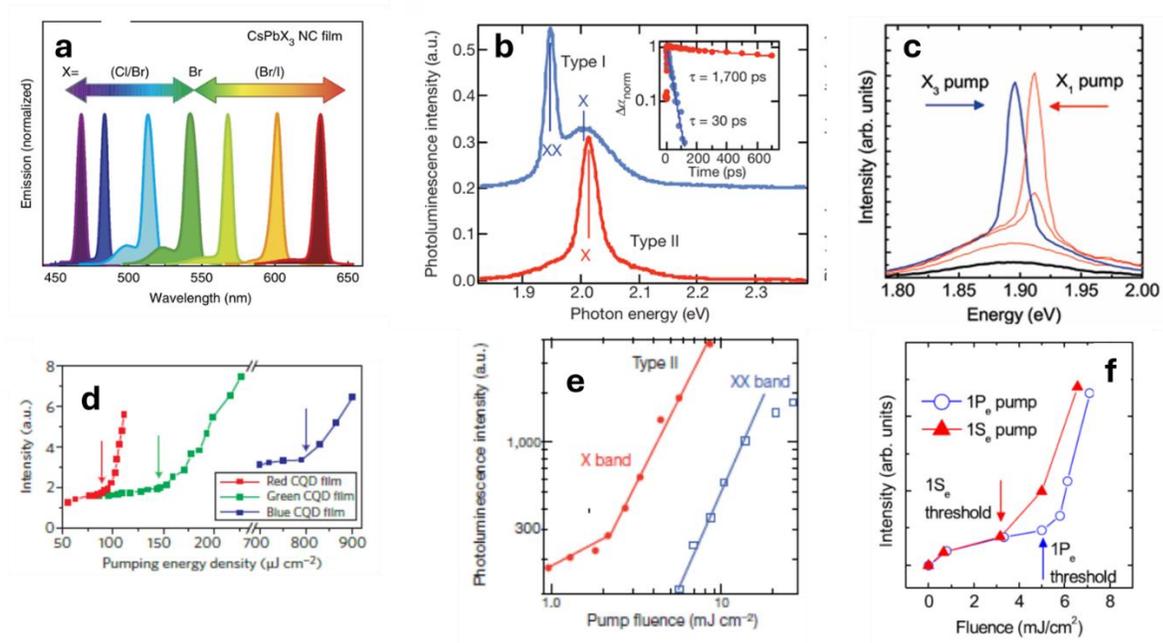

**Fig4. ASE measurements of optical gain brings measurements closer to a functioning laser.** a) ASE of lead halide perovskite CQD of different compositions to tune the spectra. Adapted from [69]. b) ASE measurements of core/shell CQD revealing gain from excitons rather than biexcitons. Adapted from [101]. c) State-resolved optical pumping in CdSe CQD shows that ASE spectra and thresholds depend upon the initially pumped exciton. Adapted from [87]. d) The buildup of ASE for different colors of CdSe CQD. Adapted from [20]. e) The buildup of ASE for different colors of inverted core/shell CQD. Adapted from [101]. f) The buildup of ASE for CdSe CQD with 1S and 1P pumping. Adapted from [87].

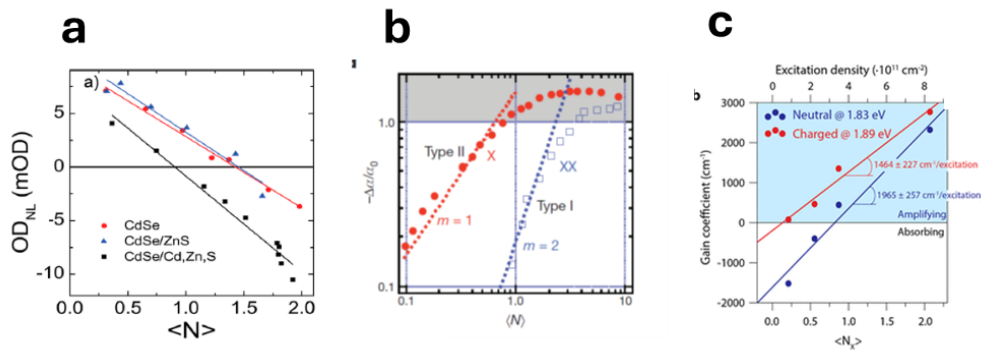

**Fig5. Optical gain thresholds are best measured by SE.** a) Thresholds for CdSe and core/shell CQD employing state-resolved optical pumping into the X1 state. Adapted from [97]. b) Inverted core/shell structures enable gain thresholds < 1. Adapted from [101]. c) Charging CQD is a recent strategy to reduce gain thresholds. Adapted from [78].

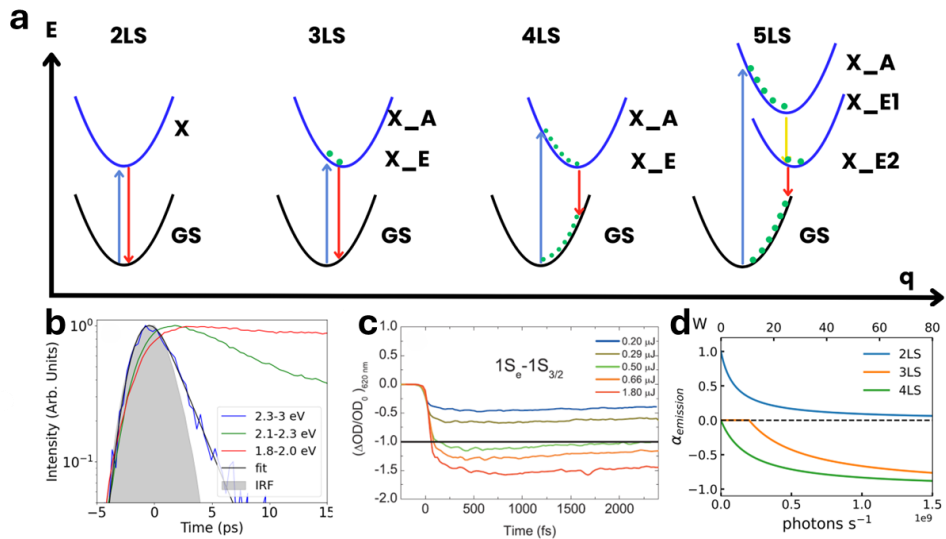

**Fig6. Evaluating CQDs as a 2 / 3 / 4+ level systems.** a) Schematic illustration of the four main level diagrams. b) Relaxation from the absorbing state to the emitting state is revealed by t-PL. Adapted from [53]. c) Buildup of SE following direct pumping into X1. Adapted from [88]. d) The behavior of classical 2 / 3 / 4 LS with continuous wave excitation.

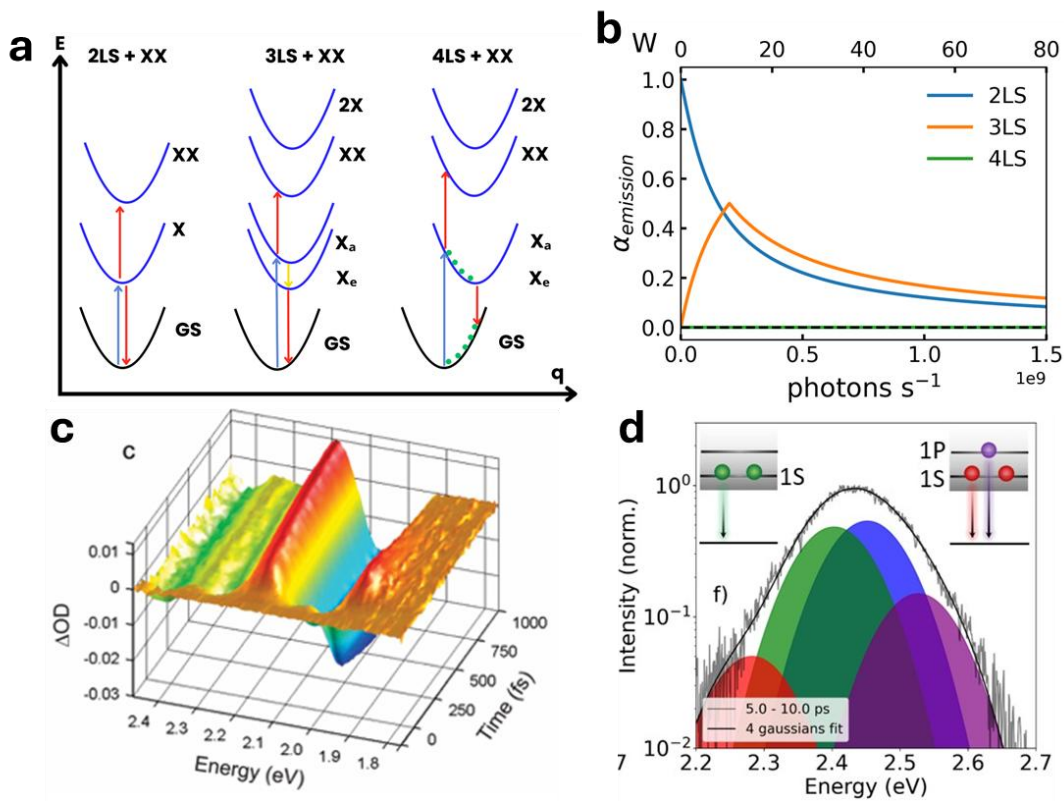

**Fig7. Adding biexcitons into the mix.** a) Schematic illustration of the three main level diagrams with addition of excited state absorption into a biexciton. Absorption from X → XX is redshifted by the biexciton binding energy. b) The kinetics of the 2 / 3 / 4 LS with excited state absorption into a biexciton. c) TA spectroscopy reveals biexciton formation. Data adapted from [50]. d) Multi-exciton emission band can be resolved by t-PL. Data adapted from [54].

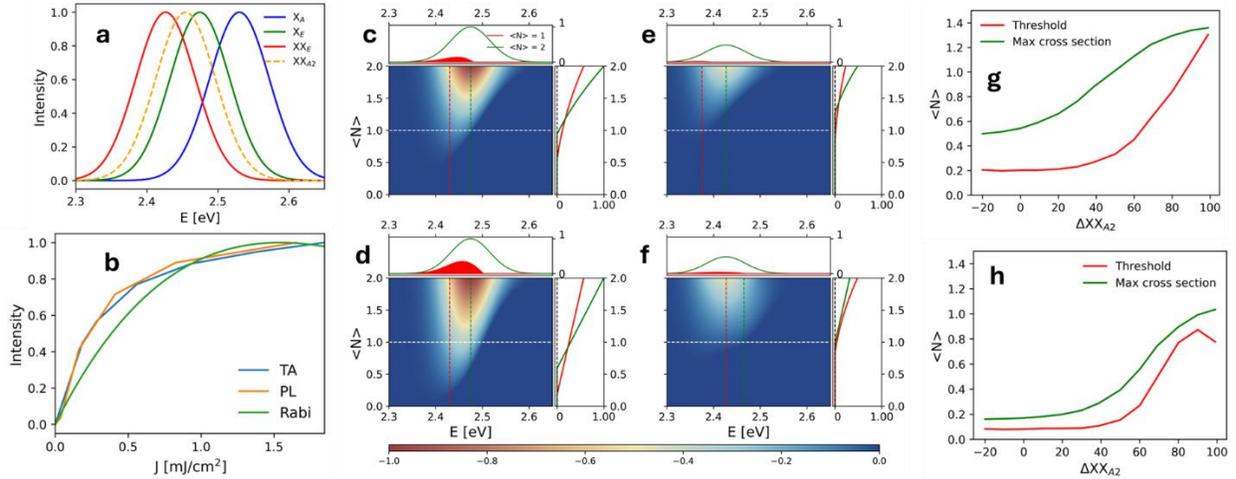

**Fig8. The Simple Model for optical gain in CQD explains and predicts the SE spectral response.**

a) The model spectra revealing the relevant transitions. b) Pumping CsPbBr3 CQD into the band edge exciton reveals saturation that is consistent with a Rabi model. c) The SE response vs <N> for a 3LS. The upper projection shows the SE spectra at two excitation densities, and the right projection shows the SE buildup vs excitation density for two probe energies. d) The same as c), but for a 4LS. e) The same as c), but with inclusion of ESA into XX. f) The same as d), but with inclusion of ESA into XX. g) The relationship between the SE threshold, <N>, and the relevant biexciton binding energy, $\Delta XX_{A2}$, for the 3LS with ESA into XX. h) The same as g), for the 4LS with ESA into XX.

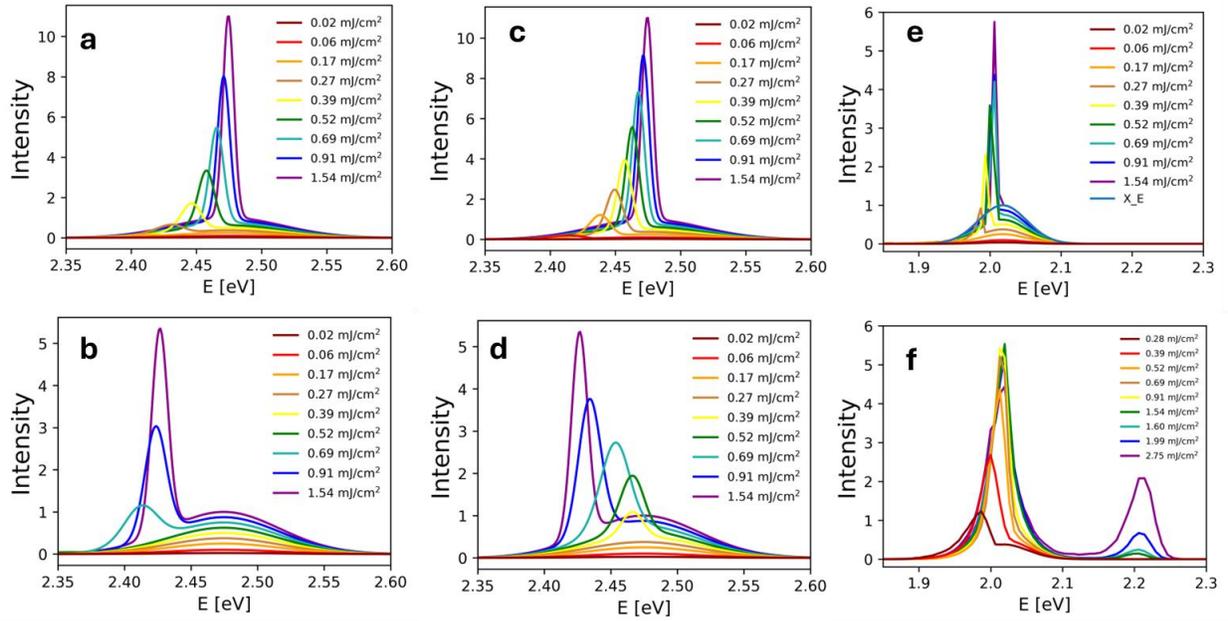

**Fig9. The Simple Model first builds the SE spectral response, which then predicts the ASE spectral response.** a) The simulated ASE spectra for a 3LS without ESA into XX. b) The simulated ASE spectra for a 3LS with ESA into XX. c) The simulated ASE spectra for a 4LS without ESA into XX. d) The simulated ASE spectra for a 4LS with ESA into XX. e) The simulated ASE spectra of the CdSe/CdS/ZnS graded alloy CQD obtained from the experimental SE spectra with $X_1$ pumping shown in Fig3e. f) The simulated ASE spectra of the CdSe/CdS/ZnS graded alloy CQD obtained from the experimental SE spectra with $X_3$ pumping shown in Fig3f.

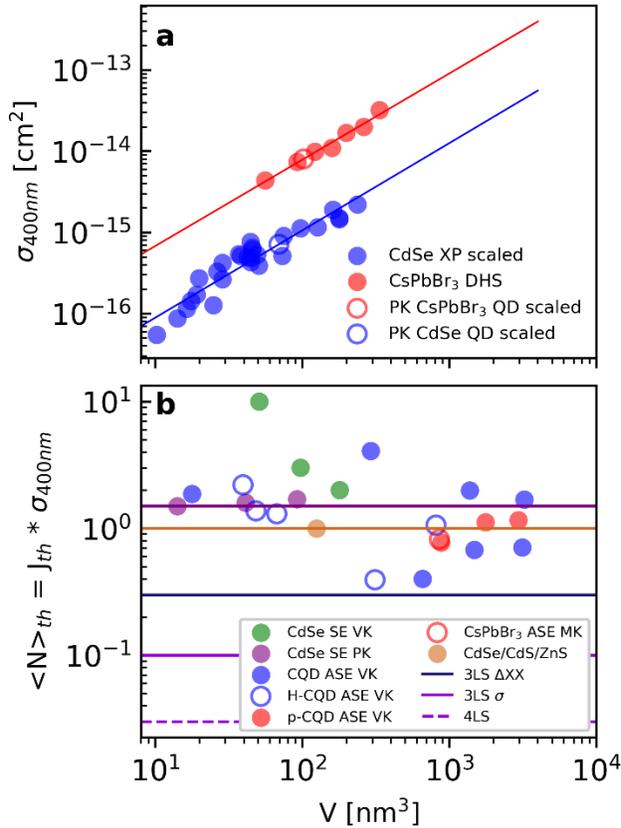

**Fig10. Identifying the true Universal ASE response for CQD, and pathways for improvement.** a) The absorption cross sections at 400 nm (3.1 eV) for CdSe and CsPbBr$_3$ CQD. CdSe data adapted from [108]. CsPbBr3 data adapted from [109]. The absolute values of each curve was obtained by identical measurements on both types of CQD for the same size [55]. b) The gain threshold in terms of photons absorbed <N> = $J_{th}\sigma_{400\,nm}$, as a function of volume shows a universal curve for CQD of two very different forms. Shown are data from Stimulated Emission (SE) and Amplified Spontaneous Emission (ASE) measurements.

**References.**

(66) Fan, F.; Voznyy, O.; Sabatini, R. P.; Bicanic, K. T.; Adachi, M. M.; McBride, J. R.; Reid, K. R.; Park, Y.-S.; Li, X.; Jain, A. Continuous-wave lasing in colloidal quantum dot solids enabled by facet-selective epitaxy. *Nature* **2017**, *544* (7648), 75-79.

(67) Fu, Y. P.; Zhu, H. M.; Stoumpos, C. C.; Ding, Q.; Wang, J.; Kanatzidis, M. G.; Zhu, X. Y.; Jin, S. Broad Wavelength Tunable Robust Lasing from Single-Crystal Nanowires of Cesium Lead Halide Perovskites (CsPbX3, X = Cl, Br, I). *ACS Nano* **2016**, *10*, 7963.

(68) Eaton, S. W.; Lai, M. L.; Gibson, N. A.; Wong, A. B.; Dou, L. T.; Ma, J.; Wang, L. W.; Leone, S. R.; Yang, P. D. Lasing in Robust Cesium Lead Halide Perovskite Nanowires. *Proc. Natl. Acad. Sci. U. S. A.* **2016**, *113*, 1993.

(69) Yakunin, S.; Protesescu, L.; Krieg, F.; Bodnarchuk, M. I.; Nedelcu, G.; Humer, M.; Luca, G. D.; Fiebig, M.; Heiss, W.; Kovalenko, M. V. Low-Threshold Amplified Spontaneous Emission and Lasing from Colloidal Nanocrystals of Caesium Lead Halide Perovskites. *Nat. Commun.* **2015**, *6*, 8056.

(70) Wang, Y.; Li, X.; Song, J.; Xiao, L.; Zeng, H.; Sun, H. All-Inorganic Colloidal Perovskite Quantum Dots: A New Class of Lasing Materials with Favorable Characteristics. *Adv. Mater.* **2015**, *27*, 7101.

(71) She, C.; Fedin, I.; Dolzhnikov, D. S.; Dahlberg, P. D.; Engel, G. S.; Schaller, R. D.; Talapin, D. V. Red, Yellow, Green, and Blue Amplified Spontaneous Emission and Lasing Using Colloidal CdSe Nanoplatelets. *ACS Nano* **2015**, *9*, 9475.

(72) Xing, G. C.; Mathews, N.; Lim, S. S.; Yantara, N.; Liu, X. F.; Sabba, D.; Gratzel, M.; Mhaisalkar, S.; Sum, T. C. Low-temperature solution-processed wavelength-tunable perovskites for lasing. *Nat. Mater.* **2014**, *13* (5), 476.

(73) Grim, J. Q.; Christodoulou, S.; Di Stasio, F.; Krahne, R.; Cingolani, R.; Manna, L.; Moreels, I. Continuous Wave Biexciton Lasing at Room Temperature using Solution Processed Quantum Wells. *Nat. Nanotechnol.* **2014**, *9*, 891.

(74) Kazes, M.; Lewis, D. Y.; Ebenstein, Y.; Mokari, T.; Banin, U. Lasing from Semiconductor Quantum Rods in a Cylindrical Microcavity. *Adv. Mater.* **2002**, *14*, 317.

(75) García-Santamaría, F.; Chen, Y.; Vela, J.; Schaller, R. D.; Hollingsworth, J. A.; Klimov, V. I. Suppressed auger recombination in "giant" nanocrystals boosts optical gain performance. *Nano Lett.* **2009**, *9* (10), 3482-3488.

(76) Malko, A.; Mikhailovsky, A.; Petruska, M.; Hollingsworth, J. A.; Klimov, V. I. Interplay between optical gain and photoinduced absorption in CdSe nanocrystals. *The Journal of Physical Chemistry B* **2004**, *108* (17), 5250-5255.

(77) Nie, Z.; Huang, Z.; Zhang, M.; Wu, B.; Wu, H.; Shi, Y.; Wu, K.; Wang, Y. Hot Phonon Bottleneck Stimulates Giant Optical Gain in Lead Halide Perovskite Quantum Dots. *ACS Photonics* **2022**, *9* (10), 3457-3465.

(78) Geuchies, J. J.; Dijkhuizen, R.; Koel, M.; Grimaldi, G.; Du Fossé, I.; Evers, W. H.; Hens, Z.; Houtepen, A. J. Zero-threshold optical gain in electrochemically doped nanoplatelets and the physics behind it. *ACS Nano* **2022**, *16* (11), 18777-18788.

(79) Qin, Z.; Zhang, C.; Chen, L.; Yu, T.; Wang, X.; Xiao, M. Electrical switching of optical gain in perovskite semiconductor nanocrystals. *Nano Lett.* **2021**, *21* (18), 7831-7838.

(80) Wang, S.; Yu, J.; Zhang, M.; Chen, D.; Li, C.; Chen, R.; Jia, G.; Rogach, A. L.; Yang, X. Stable, Strongly Emitting Cesium Lead Bromide Perovskite Nanorods with High Optical Gain